\documentclass[12pt]{article}
\usepackage{amsmath}
\usepackage{mathtools}
\usepackage{amssymb}
\usepackage{amsfonts}
\usepackage{graphicx}
\usepackage{float}
\usepackage{array}

\title{Newtonian orbits of nanoparticles interacting with structured light beams}
\date{2019}

\begin{document}
\maketitle
\subsubsection*{Authors:}
Manuel F. Ferrer-Garcia and Dorilian Lopez-Mago$^\ast$.
\subsubsection*{Addresses:}
Tecnologico de Monterrey, Escuela de Ingenier\'{i}a y Ciencias, Ave. Eugenio Garza Sada 2501, Monterrey, N.L., M\'{e}xico, 64849.\\
$^\ast$ dlopezmago@tec.mx

\section{abstract}
We perform numerical analysis to study the orbits described by subwavelength size particles interacting with structured light beams. Our solution to the particle dynamics considers: (i) the gradient force, (ii) the radiation pressure, and (iii) the force from the curl of the spin angular momentum. The last two terms, (ii) and (iii), constitute the scattering forces. The optical structures of interest are vortices, vector, and Full-Poincar\'{e} beams. From our numerical results, we show that the particle is expelled from the beam, independent of the gradient force intensity if some of the scattering forces have cylindrical symmetry. Furthermore, we found spiral orbits for some particular conditions and types of Full-Poincar\'{e} beams.


\section{Introduction}

Optical forces are usually described by two distinct components: gradient and scattering (radiation pressure) forces. The first experimental demonstration of these optical forces was  realized by Ashkin \cite{Ashkin1970}, which eventually opened the field of optical trapping and optical micromanipulation. Relevant applications include optical tweezers, particle sorting, optical levitation, along with others, mostly in the field of biophysics \cite{MacDonald2003,Lehmuskero2015,Ayala2016,DHAKAL20181}. 

Structured light beams, either in phase or polarisation, have provided new concepts and tools to control optical particles \cite{Rubinsztein-Dunlop2017}. It has been shown that the interaction of space-variant polarised light beams with nanoparticles can be described by the sum of three terms: the traditional gradient and radiation pressure forces, plus a new scattering force contribution, the spin-dependent forces, arising from the polarisation distribution \cite{albaladejo:scattering}. This opens a new degree of control in optical manipulation, i.e. we can manipulate particles via polarisation shaping. In this direction, theoretical and experimental work have been done to measure the spin-dependent forces  \cite{Angelsky2012,Bekshaev2011b}, to characterize and control the diffusion of nanoparticles \cite{Zapata2016,Albaladejo2009a}, and to understand complex particle dynamics \cite{Berry2016}. 

Berry and Shukla have shown that these scattering forces are physical examples of curl forces, which present interesting Newtonian dynamics \cite{Berry2012,Berry2013,Berry2015,Berry2016}. They showed that if the curl forces present rotational symmetry, the particles experience an outward spiral motion and escape the gradient force, regardless of its strength. On the contrary, curl forces without rotational symmetry showed chaotic behaviour. These findings have implications in current and future experiments with trapped particles in vacuum \cite{Reimann2018}.

The question arises if we can observe these findings using well known structured optical light beams, such as Laguerre-Gauss beams \cite{Allen1992}, vector beams \cite{Zhan2009}, and Full-Poincar\'{e} beams \cite{alonso:FPB}. Since these structures are routinely realized in the laboratory, it is of interest to study the dynamics due to their optical forces. Therefore, in this work, we study the Newtonian dynamics of sub-wavelength size particles interacting with these optical beams. 

The structure of the paper is as follows. Section 2 presents the mathematical model and explains the numerical method used in this work to solve for the particle dynamics. Section 3 analyses the forces arising from a single vortex beam with uniform polarisation (specifically, a Laguerre-Gauss beam). The last sections are dedicated to explore unconventional polarisation distributions, for example, section 4 studies vector and Full-Poincar\'{e} patterns, which are generated by the collinear superposition of two optical vortices, while section 5 explores the superposition of two off-axis optical vortices.

\section{Mathematical and numerical method}

First, we establish notation and briefly explain the dynamical system of our interest. The optical forces acting on a subwavelength-sized particle, illuminated with a monochromatic light beam with electromagnetic fields $\mathbf{E}$ and $\mathbf{H}$, are modeled under the dipole approximation \cite{albaladejo:scattering}, which results in the following expression: 
\begin{equation}
\langle \mathbf{F} \rangle= \frac{1}{4} \mbox{Re}\left[\alpha \right] \nabla  | \mathbf{E} |^2 + \frac{\sigma}{2c} \mbox{Re}\left[\mathbf{E} \times\mathbf{H^*}\right]  +\frac{\sigma}{2} \frac{\epsilon_0}{k}\mbox{Re}\left[i(\mathbf{E} \cdot \nabla ) \mathbf{E^*}\right].
\label{eq:forceexp}
\end{equation}
In this result, $\langle \mathbf{F} \rangle$ stands for the time-averaged force,   $k=\omega/c=2\pi/\lambda$, where $\omega$ and $\lambda$ are the angular frequency and vacuum wavelength of the electromagnetic field, respectively, $\sigma=k \mbox{Im}[\alpha]/\epsilon_0$ is the total cross-section of the particle and $\alpha$ its complex polarizability. For a particle of size $a$ (with $a<<\lambda$) and with relative permittivity $\epsilon$, 
\begin{equation}
\alpha=\frac{\alpha_0}{1-i\alpha_0 k^3 /(6\pi\epsilon_0)}, \qquad \mathrm{with} \qquad \alpha_0=4 \pi \epsilon_0 a^3 \frac{\epsilon-1}{\epsilon+2}.
\end{equation}
We can identify three terms in equation \ref{eq:forceexp}: the intensity-dependent gradient force $\mathbf{F}_\mathrm{G} \propto \nabla |\mathbf{E}|^2$, the transverse radiation pressure contribution $\mathbf{F}_\mathrm{RP}\propto \mathrm{Re}[\mathbf{E} \times \mathbf{H^{\ast}}]$, and a force that is modulated by the polarisation distribution, which is called the curl force $\mathbf{F}_\mathrm{C} \propto \mathrm{Re} [i(\mathbf{E}\cdot \nabla )\mathbf{E}^{\ast}]$. This third term is usually neglected for most applications due the assumption of either perfect plane waves with a uniform polarisation distribution or the use of particles whose radius is bigger that the wavelength. Nevertheless, trapping and complex trajectories due to curl forces near optical vortices have been studied for simpler models, where the particles follow spiral paths before being expelled out of the beams  \cite{Berry2013}. In contrast to the gradient force, the two other terms of equation \ref{eq:forceexp} are non-conservative and therefore, the dynamics of the particle cannot be described by Hamiltonian mechanics, and hence, they are governed by Newton's second law of motion \cite{Berry2016}. Thus, a nanoparticle illuminated by an optical field describes a trajectory $\mathbf{r}(t)$ due the exerted force, where $\mathbf{r}(t)$ is found by solving Newton's second law
\begin{equation}
m \frac{ d^2 \mathbf{r}}{dt^2}= \langle \mathbf{F} \rangle.
\label{eq:difforce}
\end{equation}

We have neglected the existence of other forces due to the interaction of the particle with a medium or thermodynamic effects, such as the Brownian motion. Analytical solutions to equation \ref{eq:difforce} are difficult or even impossible to obtain, so we have performed numerical simulations were the optical forces on every position $\mathbf{r}(t)$ were calculated and introduced into a Verlet algorithm and validated with a fourth-order Runge-Kutta method using a computational toolbox developed in our previous work \cite{ferrer:compu}.

For the description of the electromagnetic fields, we consider paraxial laser beams propagating along the $z$ axis of a coordinate system $\mathbf{r}=(\mathbf{r}_t , z)$ , where $\mathbf{r}_t = (x \mathbf{\hat{x}} + y \mathbf{\hat{y}} )$ is the transverse radius vector. The fields are written as $\mathbf{E} = (\mathbf{E}_t + \mathbf{\hat{z}} E_z)\exp(ikz)$ and $\mathbf{H} = (\mathbf{H}_t + \mathbf{\hat{z}} H_z)\exp(ikz)$, where the subscripts $t$ and $z$ stand for transverse and longitudinal components, respectively. Notice that an adequate description of the radiation pressure forces requires the longitudinal components of $\mathbf{E}$ and $\mathbf{H}$ \cite{carnicer:longi}. In other words, if the electromagnetic field lies in the transverse plane, the resulting radiation-pressure force is directed along $z$, and the transverse component (responsible for the orbital motion of particles) is thus ignored. Therefore, we obtain the $z$ component of $\mathbf{E}$ from the perturbative series expansion of Maxwell equations provided by Lax \textit{et al.} \cite{Lax1975}, where the longitudinal fields are calculated from the transverse components through  
\begin{equation} \label{Eq:longitudinal}
\{ E_z , H_z \} = \frac{i}{k}  \nabla_t \cdot \{ \mathbf{E}_t , \mathbf{H}_t \},
\end{equation}
where $\nabla_t= \mathbf{\hat{x}} \partial_x+\mathbf{\hat{y}} \partial_y$ is the transverse nabla operator. 

The light beams of our interest are constituted by superpositions of optical vortices. We make use monochromatic vortices described by single-ringed Laguerre-Gauss modes (\textit{i.e.} with radial index $n=0$) and characterized by a helical phase of topological charge $\ell=0,\pm1,\pm2,\ldots$. Their complex field amplitude, normalized to unit power (\textit{i.e.} $\int \int |U|^2 \mathrm{d}x \mathrm{d}y=1$ ) and propagating along the $z$ axis is written as  
\begin{equation}
\mbox{U}_\ell(x,y;z=0)=C_\ell \, (r/w_0)^{|\ell|} \, \exp(-r^2/w_0^2) \, \exp(i \ell\phi),
\label{LGB}
\end{equation}
where $(x,y)=(r\cos \phi , r\sin \phi )$,  $\omega_0$ if the waist of the Gaussian envelope, and $C_\ell = (2^{|\ell|+1}/w_0^2 \pi |\ell|!)^{1/2}$ is the normalisation constant. In order to generate a space-dependent polarisation structure, we consider a transverse electric field $\mathbf{E}_t$ formed by the coherent superposition of two optical vortices of the form
\begin{equation}
    \mathbf{E}_t(x,y)= \cos \eta  U_{\ell_1}(x+x_0,y) \mathbf{\hat{e}_1} + e^{i\delta}\sin \eta U_{\ell_2}(x-x_0,y) \mathbf{\hat{e}_2}.
    \label{eq:modelop}
\end{equation}

\begin{figure}[H]
     \centering
    \includegraphics[width=12 cm]{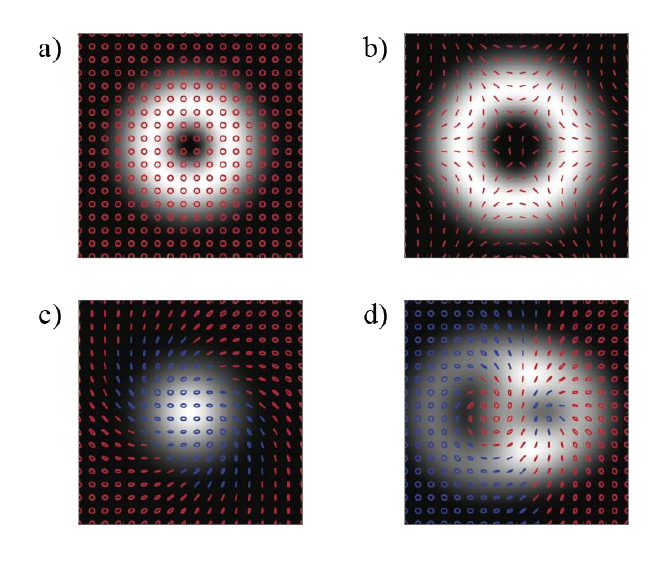}
     \caption{Intensity and polarisation distributions for some possible configuration of the beam described in equation \ref{eq:modelop} under different parameters values (a) $\mathbf{C}=(0, \, \pi/4, \, \pi/2, \, 0, \, 1,  \, \ell_2,  \, 0)$, (b) $\mathbf{C}=(\pi/4, \, \pi/4, \, \pi/2, \, 0, \,  2,  \, 2,  \, 0)$, (c) $\mathbf{C}=(\pi/8, \, \pi/6, \, -\pi/2, \, \pi/2, \,  0,  \, -2,  \, 0)$  and (d) $\mathbf{C}=(\pi/4, \, \pi/4, \, -\pi/2, \, \pi/4, \,  2,  \, -1,  \, 0.5)$.}
     \label{fig:exfield}
 \end{figure}

Variations in the relative amplitude and phase between both beams are allowed by the parameters $\eta$ and $\delta$, while the polarisation structured also depends on the separation distance between the centroid of the beams $\Delta x =2x_0$. We also consider an elliptical polarisation basis $\mathbf{\hat{e}_1}$ and $\mathbf{\hat{e}_2}$, which are complex and orthogonal unitary vectors of the form
\begin{eqnarray}\label{Eq:ellipol}
    \mathbf{\hat{e}_1}=\cos \theta \, \mathbf{\hat{x}}+e^{i\beta}\sin \theta \, \mathbf{\hat{y}}, \\
    \mathbf{\hat{e}_2}=\sin \theta \, \mathbf{\hat{x}}-e^{i\beta}\cos \theta \, \mathbf{\hat{y}},
\end{eqnarray}
where $\theta$ (measured with respect to $+x$) determines the relative amplitude between their Cartesian components and $\beta$ controls the relative phase. For the sake of a simpler notation, a vector $\mathbf{C}=(\eta, \, \theta, \, \beta, \, \delta, \,  \ell_1,  \, \ell_2,  \, x_0/w_0)$ is defined so that it contains all the parameters for a specific configuration. Examples of possible structured beams generated by equation \ref{eq:modelop} are depicted  in figure \ref{fig:exfield} while table \ref{tab:dic} contains the value of the parameters for the most common vector beams. 

\begin{table}[t]
\centering
\begin{tabular}{ | m{5.5cm} | c | c | c | c | c | c | c | }
\hline
 Description & $\eta$  & $\theta$ & $\beta$ & $\delta$ & $ \ell_1$ & $\ell_2$ & $x_0/w_0$ \\ \hline
  Optical vortex with topological charge $\ell$ and linearly polarised with an inclination angle of $\theta$ & 0 & $\theta$ & 0 & - & $\ell$ & - & 0 \\ \hline
   Optical vortex with topological charge $\ell$ and circular polarisation $\hat{\mathbf{c}}_{\pm}$ & 0 & $\pm \pi/4$ & $\pi/2$ & - & $\ell$ & - & 0 \\ \hline
    Radially polarised vector beam & $\pi/4$ & $\pi/4$ & $\pi/2$ & 0 & -1 & 1 & 0 \\ \hline
     Azimuthally polarised vector beam & $\pi/4$ & $\pi/4$ & $\pi/2$ & 0 & 1 & -1 & 1 \\ \hline
      Full Poincar\'{e} beam carrying a lemon singularity & $\pi/4$ & $\pi/4$ & $\pi/2$ & 0 & 0 & 1 & 0 \\ \hline
       Full Poincar\'{e} beam carrying a star singularity & $\pi/4$ & $\pi/4$ & $\pi/2$ & 0 & 1 & 0 & 0 \\ \hline
\end{tabular}
\caption{Values of the parameters in equations \ref{eq:modelop} and \ref{Eq:ellipol}  to generate standard structured beams.}
\label{tab:dic}
\end{table}

\section{Optical forces produced by a single optical vortex}

As first approach, let us consider a single optical vortex with uniform polarisation distribution $\mathbf{E}_{S,t}= U_\ell \, \mathbf{\hat{e_1}}$
where $U_\ell$ and $\mathbf{\hat{e_1}}$ are given by equations \ref{LGB} and \ref{Eq:ellipol}, respectively. Then, we determine the longitudinal component $\mathbf{E}_{S,z}$ using equation \ref{Eq:longitudinal}. Substituting $\mathbf{E}_S= \mathbf{E}_{S,t} + \mathbf{E}_{S,z}$  in equation \ref{eq:forceexp}, we obtain the following expressions for the transverse optical forces:

\begin{align}
\mathbf{F}_{\mathrm{G}} = & \frac{\mbox{Re}[\alpha] \ |\mbox{U}_\ell|^2 }{2} \left( \frac{|\ell|}{r}-\frac{2r}{w_0^2} \right) \, \, \mathbf{\hat{r}}, \label{Eq:gradforc} \\
\mathbf{F}_{\mathrm{RP}}=& \frac{\mbox{Im}[\alpha]|\mbox{U}_\ell|^2 }{2} \left[ \frac{\ell}{r}- \sin(2\theta) \sin(\beta) \left(\frac{|\ell|}{r}-\frac{2r}{w_0^2}\right) \right]  \mathbf{\hat{\boldsymbol{\phi}}}, \label{Eq:EqRP}\\
\mathbf{F}_{\mathrm{C}} = &\frac{\mbox{Im}[\alpha]|\mbox{U}_\ell|^2 }{2} \left[ \frac{\ell \cos^2\theta \sin\phi }{r}+ \frac{\sin(2\theta) \sin \beta \sin\phi  }{2}\left(\frac{|\ell|}{r} -\frac{2r}{w_0^2}\right) \right. \nonumber \\
     & \left. -\frac{\ell \sin(2\theta)\cos\beta\cos\phi}{2r}\right] \mathbf{\hat{x}} +\frac{\mbox{Im}[\alpha]|\mbox{U}_\ell|^2 }{2} \left[ -\frac{\ell \sin^2\theta\cos \phi }{r} \right. \nonumber \\
     &  \left.- \frac{\sin(2\theta)  \sin \beta \cos \phi }{2}\left(\frac{|\ell|}{r}-\frac{2r}{w_0^2}\right) + \frac{\ell \sin(2\theta) \cos \beta \sin \phi}{2r}\right] \mathbf{\hat{y}}\label{Eq:Fc}.
     \end{align}
These equations constitute one of the first important results of this work. It seems that such expressions should be well-known, but so far, we have not been able to find it in the literature. The importance of these expressions, is that they show that the scattering forces ($\mathbf{F}_\mathrm{RP} + \mathbf{F}_\mathrm{C}$), and in particular the curl force contribution, are always present in homogeneously-polarised LG beams, except for the particular case when $\beta=\ell=0$, which corresponds to a fundamental Gaussian beam with uniform linear polarisation. Scattering forces can generate undesirable effects, such as chaotic dynamics and trapping instabilities \cite{novotny:cancel}.

As expected, the gradient force is independent of polarisation (notice the absent of $\beta$ and $\theta$ in equation \ref{Eq:gradforc}), presents rotational symmetry, and has radial direction. The angular index determines the inflection point of the force  (\textit{i.e.} when it changes direction). It is straightforward to show that the force is pointing towards the maximum intensity, located at $r_{max}^{2}=|\ell| w_{0}^{2}/2$. Therefore, the particle should be trapped or oscillate around the maximum intensity of the ring. On the contrary, the scattering forces depend on polarisation as well as in the angular index $\ell$. For example, when we consider linear polarisation ($\beta=0$), the expressions for the scattering forces become 
\begin{align}
\mathbf{F_\mathrm{RP}} &= \frac{\ell|\, \mbox{Im}[\alpha]|\mbox{U}_\ell|^2 }{2r} \mathbf{\hat{\boldsymbol{\phi}}}, \\[1.5ex]
    \mathbf{F}_\mathrm{C} &= \frac{\ell|\, \mbox{Im}[\alpha]|\mbox{U}_\ell|^2 }{2r} \left[ \left(\cos^2\theta \sin\phi - \frac{\sin(2\theta)\cos\phi}{2} \right)\mathbf{\hat{x}}\right. \nonumber \\[1.5ex]
     & \left. + \left(-\sin^2\theta \cos\phi + \frac{\sin(2\theta)\sin\phi}{2} \right)\mathbf{\hat{y}}\right],
\end{align}
while for circular polarisation ($\theta=\pi/4, \beta=\pm \pi/2$)
\begin{align}
\mathbf{F}_\mathrm{RP} & = \frac{\mbox{Im}[\alpha]|\mbox{U}_\ell|^2 }{2} \left[ \frac{\ell}{r} \mp  \left(\frac{|\ell|}{r}-\frac{2r}{w_0^2}\right) \right] \mathbf{\hat{\boldsymbol{\phi}}}, \\[1.5ex]
\mathbf{F}_\mathrm{C} & = \frac{\mbox{Im}[\alpha]|\mbox{U}_\ell|^2 }{2} \left[ \frac{\ell  }{r}\pm \left(\frac{|\ell|}{r} -\frac{2r}{w_0^2}\right)\right] \mathbf{\hat{\boldsymbol{\phi}}}.
\end{align} 

\begin{figure}[H]
     \centering
    \includegraphics[width=12 cm]{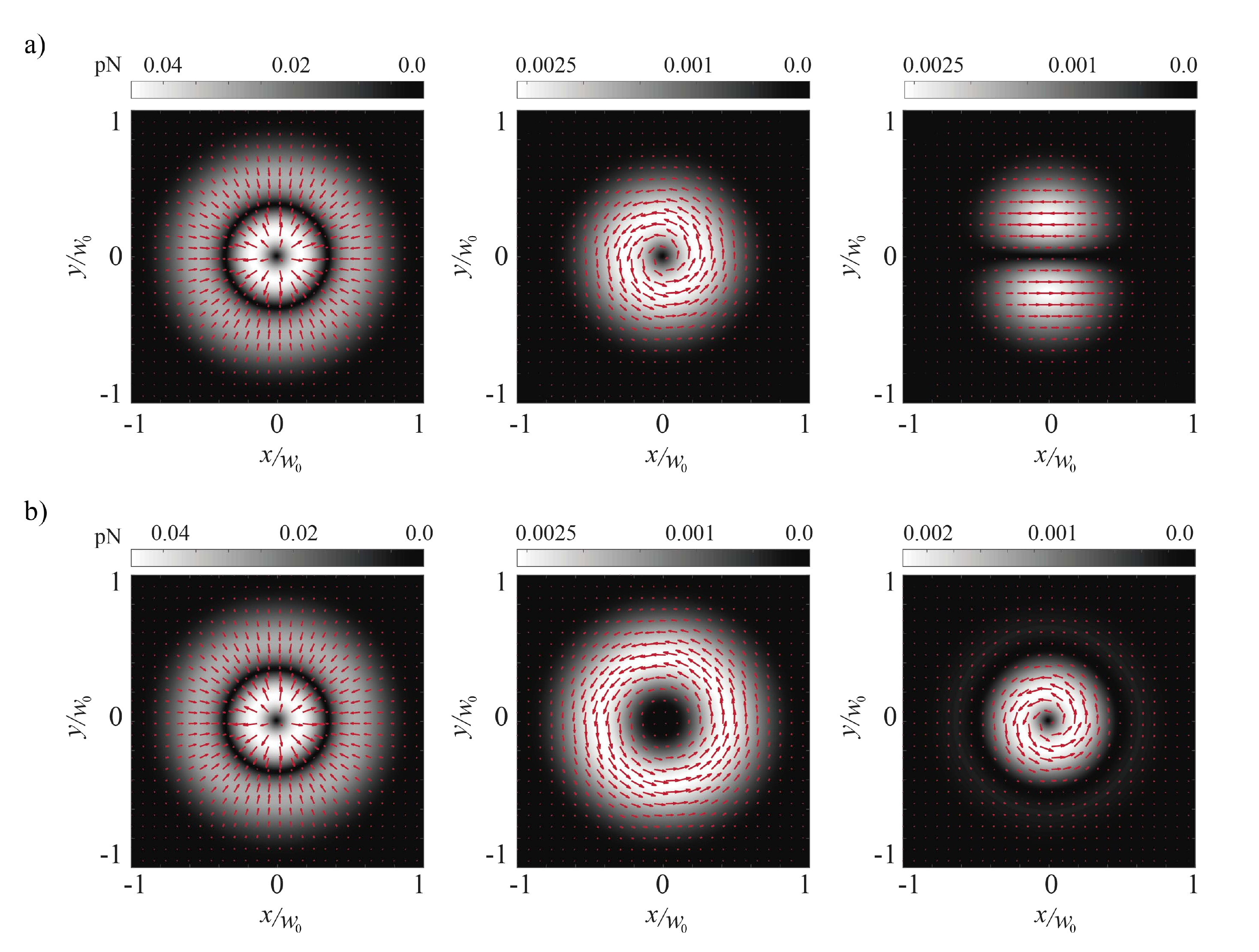}
     \caption{Optical forces for a single-ringed Laguerre-Gauss with $\ell = 1$ under initial conditions $x_0/w_0=0.3,y_0/w_0=0,v_{x_0}=0,v_{y_0}=0$ for (a) horizontal polarisation  $\mathbf{C}=(0,\, 0,\,0,\,0,\,1,\,\ell_2,\,0)$ and (b) circular polarisation $\mathbf{C}=(0,\, \pi/4,\,\pi/2,\,0,\,1,\,\ell_2,\,0)$. The first, second and third columns show the gradient, radiation pressure and curl forces, respectively, given by equations \eqref{Eq:gradforc} to \eqref{Eq:Fc}. The colormap indicates the force amplitude whereas the arrows indicate direction.}
     \label{fig:ForcesSingle}
 \end{figure}

Figure \ref{fig:ForcesSingle} shows the three components of the optical forces produced by a Laguerre-Gauss beam with $\ell=1$ and having linear or circular polarisation. In both cases the gradient force is identical and the radiation pressure shows azimuthal symmetry. However, notice that for circular polarisation, the radiation pressure shows a spin-dependent contribution. Depending on the relative handedness between the orbital angular momentum (given by the sign of $\ell$, positive $\ell$ is left-handed and negative $\ell$ is right-handed) and polarisation (given by the sign of $\beta$, where positive $\beta$ is left-handed and negative $\beta$ is right-handed), the radiation pressure and the curl force for circular polarisation can present one or two rings of forces. One ring occurs for opposite handedness whereas two rings appear for equal handedness. 
 
\begin{figure}[ht]
     \centering
     \includegraphics[width=12 cm]{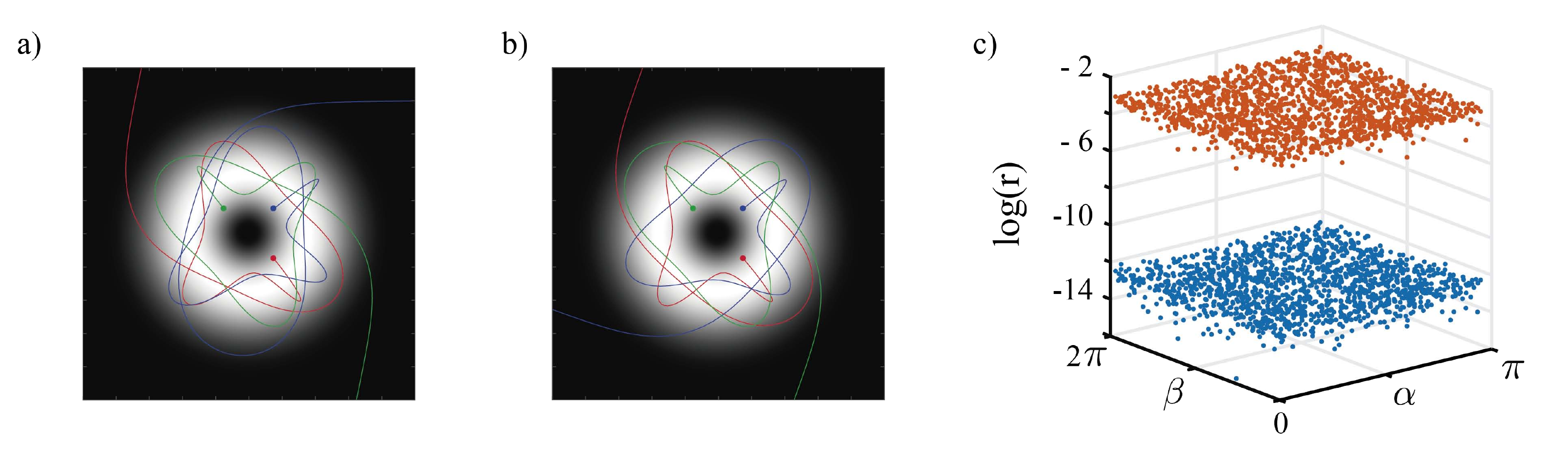}
     \caption{ Free-space dynamics of a particle interacting with Laguerre-Gauss beam with $\ell=1$ for (a) horizontal polarisation $\mathbf{C}=(0,\, 0,\,0,\,0,\,1,\,\ell_2,\,0)$ and (b) circular polarisation $\mathbf{C}=(0,\, \pi/4,\,\pi/2,\,0,\,1,\,\ell_2,\,0)$. (c) Initial (blue) and final (orange) particle distance for different polarisation states characterised by the parameters $\theta$ and $\beta$ after $t=0.1$.}
     \label{fig:DifferentInitialconditions}
 \end{figure}

 The dynamic of a particle interacting with a Laguerre-Gauss beam is shown in figure \ref{fig:DifferentInitialconditions}. Figures \ref{fig:DifferentInitialconditions}(a) and \ref{fig:DifferentInitialconditions}(b) show the results for linear and circular polarisations, respectively. We present the trajectories of three identical particles located at different angular positions. We observe that the nanoparticle dynamics preserve the rotational symmetry of the optical forces. For the linear case, the dynamic has a two-fold rotational symmetry with respect to the inclination angle of the polarisation state, whereas for the circular polarisation there is a full rotational symmetry.
 
 Furthermore, figure \ref{fig:DifferentInitialconditions}(c) shows an exhaustive numerical analysis that shows that independently of the polarisation state of the vortex and the initial condition for the particle, the resultant trajectory is a complex open orbit in which the particle is expelled due to the non-conservative scattering force components and despite the action of the gradient force to trap the particle. This result agrees to the conclusions given by Berry et al. \cite{Berry2013}, where it is shown that despite the intensity of the gradient force, a curl force presenting cylindrical symmetry will always expel the particle.
  
\section{Optical forces produced by an on-axis superposition of optical vortices}
\label{Sec:on-axis}
\subsection{Cylindrical Vector Beams}

\begin{figure}[ht]
     \centering
     \includegraphics[width=12 cm]{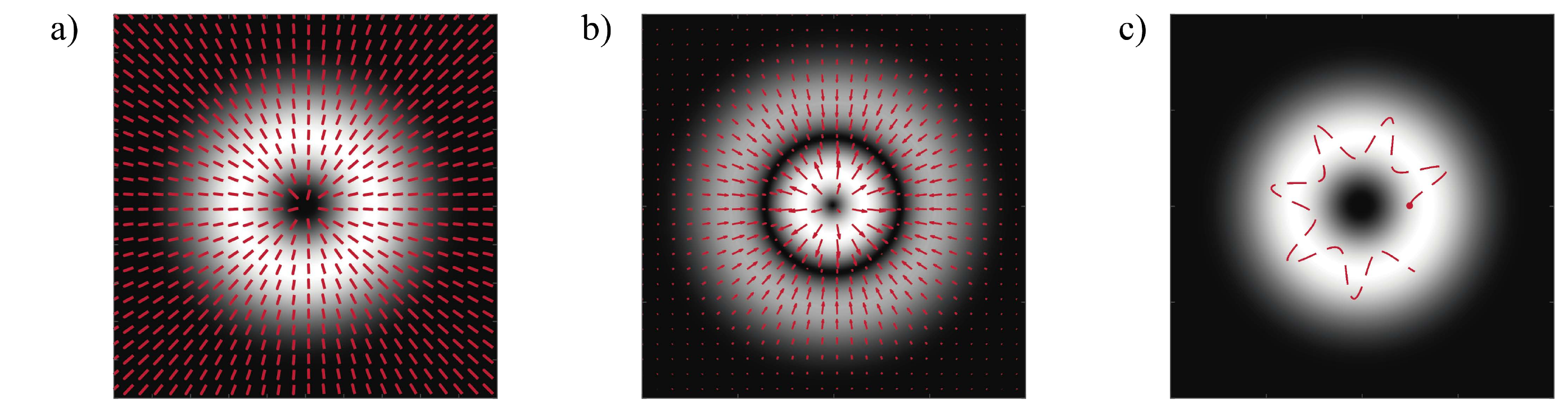}
     \caption{Radially polarised vector beam: a) Intensity and polarisation structure b) Gradient force c) Simulated trajectory of a particle with a constant initial velocity in the $y$ direction.}
     \label{fig:CVexp}
\end{figure}

Cylindrical vector (CV) beams have a single-ringed intensity pattern similar to a single optical vortex, but they have an inhomogeneous polarisation distribution. Therefore, the comparison between CV beams and optical vortices serves to analyse polarisation effects on the nanoparticle dynamics. The CV beam family is written as \cite{Zhan2009},
\begin{equation}
   \mathbf{E}_{CV,t}(x,y)=\frac{1}{\sqrt{2}}\, U_{\ell}(x,y) \, \mathbf{\hat{c}_+} +  \frac{e^{i\delta}} {\sqrt{2}}\, U_{-\ell}(x,y) \, \mathbf{\hat{c}_-},  \label{eq:CVB}
\end{equation}
where $\ell \neq 0$ and $\mathbf{\hat{c}}_{\pm} = \mathbf{\hat{x}} \pm i\mathbf{\hat{y}}$ are the circular polarisation basis. For our purposes, it is convenient to write this expression using a linear polarisation basis as
\begin{align}
   \mathbf{E}_{CV,t}(x,y)=\frac{C_\ell}{\sqrt{2}C_0}\, \left(\frac{r}{w_0}\right)^{|\ell|}\,U_0
   &\left[   \tau_{\mathbf{\hat{x}}} \, \mathbf{\hat{x}} +\tau_{\mathbf{\hat{y}}} \,  \mathbf{\hat{y}} \right],
   \label{eq:CVB1}
\end{align}
where $\tau_{\mathbf{\hat{x}}}=(1+e^{i\delta})\cos\ell\phi+i(1-e^{i\delta})\sin\ell\phi$, $\tau_{\mathbf{\hat{y}}}=i(1-e^{i\delta})\cos\ell\phi+(1+e^{i\delta})\sin\ell\phi$. 
Substituting the latter expression into equation \ref{eq:forceexp}, we observe that the gradient force expression is equal to that obtained for the case of the corresponding single optical vortex since $|\mathbf{E_{S,t}}|^2=|\mathbf{E}_{CV,t}|^2$, while for the scattering terms, the result is not so straightforward. After some algebra we find that 
\begin{align}
    \mathbf{F}_\mathrm{RP}\propto & \text{Re}\left[i \, \tau \nabla_t \left[ \left(\frac{r}{w_0}\right)^{|\ell|}\,U_0 \right] \right]=0,\\[1.5ex]
    \mathbf{F}_C \propto & \text{Re} \left[ i \sum_{\mathbf{\hat{v}}=\{\mathbf{\hat{x}},\mathbf{\hat{y}}\}}\left( \tau_{\mathbf{\hat{v}}}^* \tau_\mathbf{\hat{x}} \partial_x + \tau_{\mathbf{\hat{v}}}^* \tau_\mathbf{\hat{y}} \partial_y  \right) \left[ \left(\frac{r}{w_0}\right)^{|\ell|}\,U_0 \right] \mathbf{\hat{v}} \right]=0, \label{bubu}
\end{align}
where $\tau=|\tau_\mathbf{\hat{x}}|^2+|\tau_\mathbf{\hat{y}}|^2$,  $\tau_j^* \tau_k=2 \sin(\delta - 2 \ell \phi)$ for $j\neq k$ and $ \nabla U_0 \in\mathbb{R}$. In both of the previous expressions, the terms inside the brackets are always purely imaginary, implying that scattering forces in CV beams vanish independently on the topological charge and the relative phase $\delta$ between the circular components of the field. This result coincides with the numerical results shown by Ferrer et al. \cite{ferrer:compu} and represents the second important result of the present work.  Since only the gradient force is not zero, the particle oscillates with respect to the closest intensity maximum describing an stable orbit as shown in figure \ref{fig:CVexp}. 

\subsection{Full Poincar\'{e}}

\begin{figure}[ht]
\centering
\includegraphics[width=10 cm]{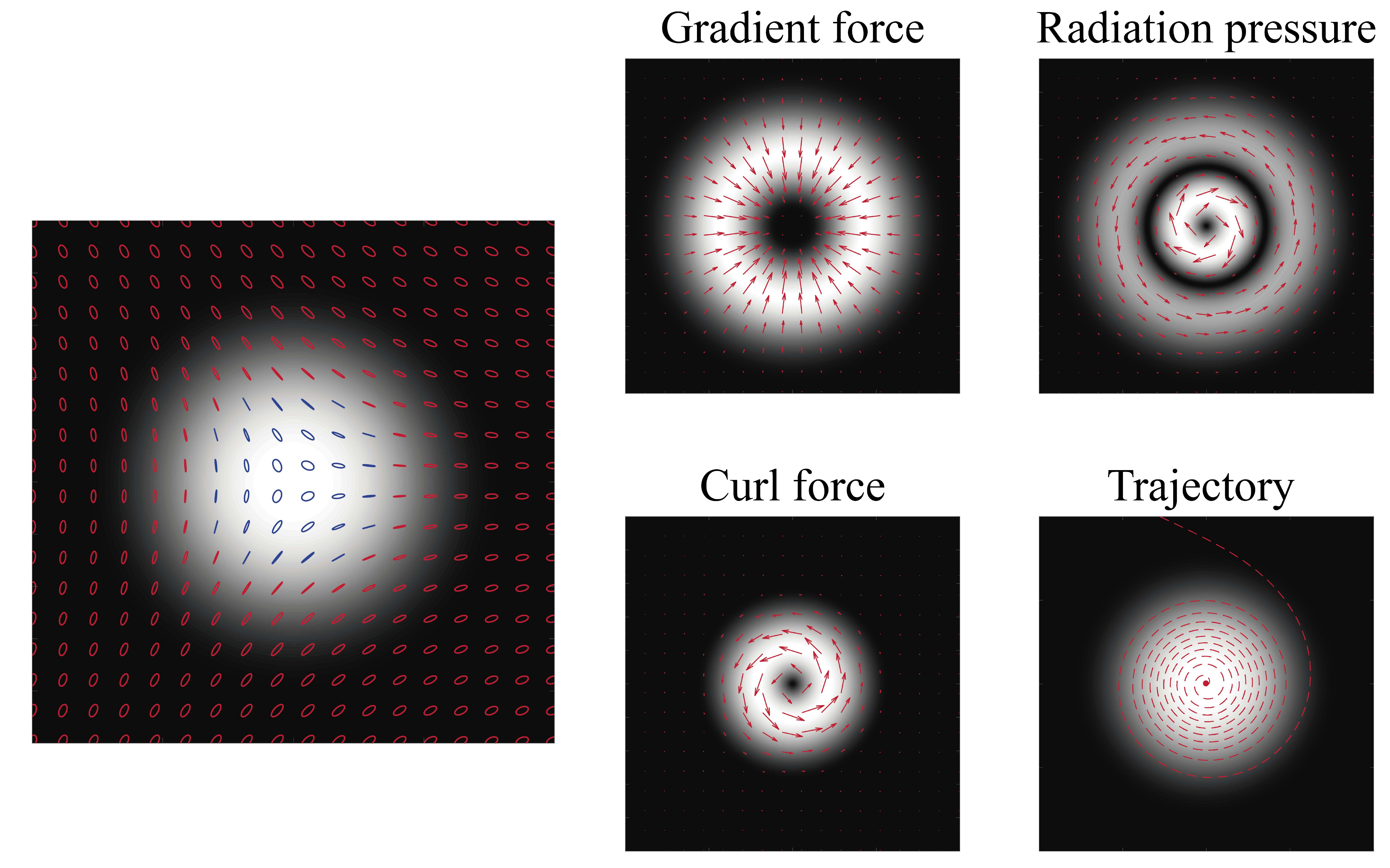}
\caption{Intensity and polarisation structure, curl force and trajectory for a Full Poincar\'e beam carrying a star singularity.}
\label{fig:FPBexp1}
\end{figure}

\begin{figure}[ht]
\centering
\includegraphics[width=10 cm]{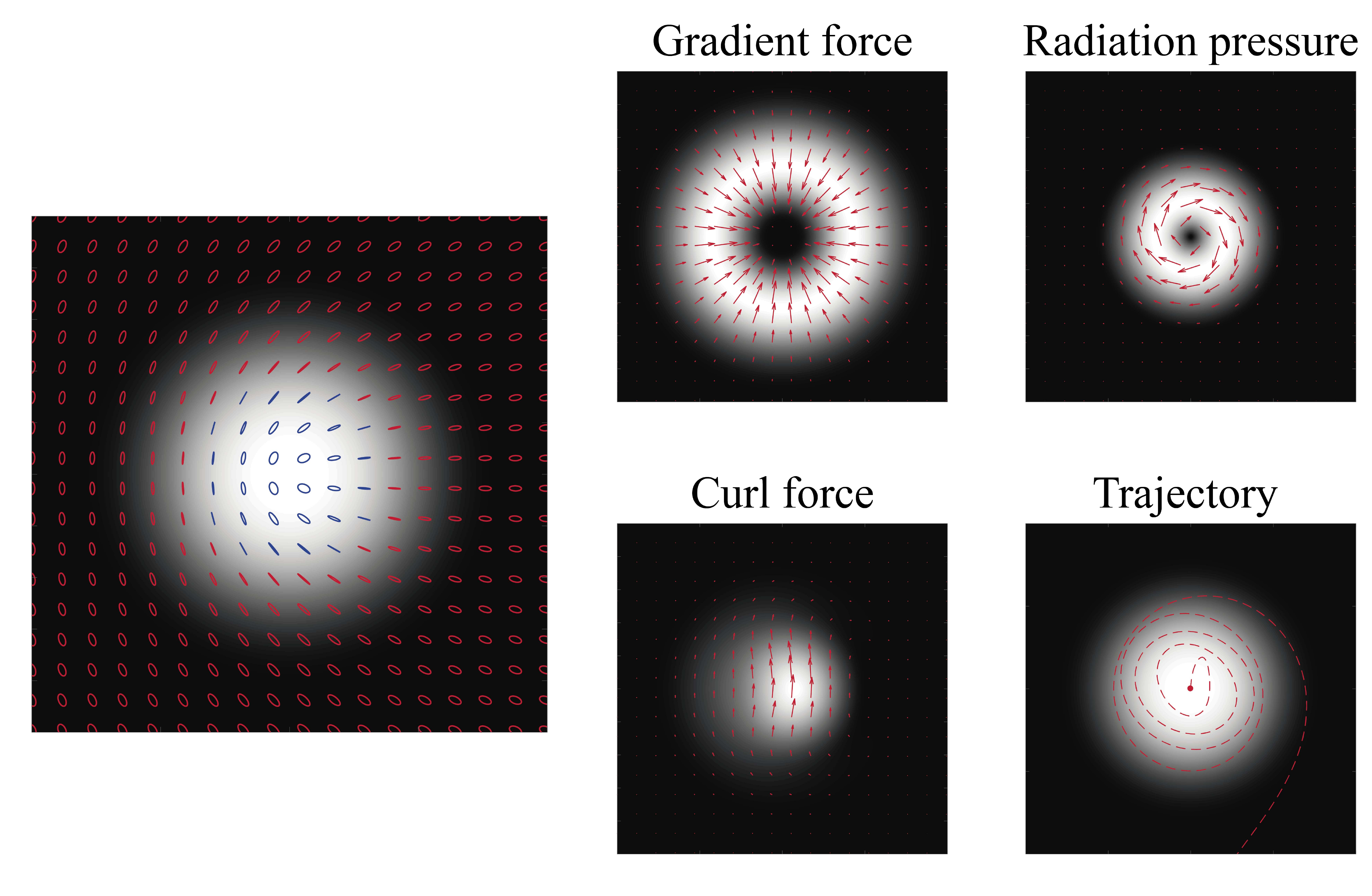}
\caption{Intensity and polarisation structure, curl force and trajectory for a Full Poincar\'e beam carrying a lemon singularity.}
\label{fig:FPBexp2}
\end{figure}

Another common vector beams are the Full Poincar\'{e} (FP) beams  \cite{alonso:FPB} in which each point represents a polarisation state on the Poincar\'{e} sphere and it can be written in terms of the circular basis as 
\begin{equation}
\mathbf{U}(x,y,z=0)= \frac{1}{\sqrt{2}} \, U_{0}(x,y)\,  \mathbf{\hat{c}_h} +\frac{1}{\sqrt{2}} \, U_{1}(x,y)\, \mathbf{\hat{c}_{\bar{h}}}
\label{eq:FPB1}
\end{equation}
 where $\mathbf{\hat{c}_h}$ stands for $\mathbf{\hat{c}_+}$ or $\mathbf{\hat{c}_-}$ and $\mathbf{\hat{c}_{\bar{h}}}$ for its respective orthogonal state. Depending on the handedness of $\mathbf{\hat{c}_h}$, the polarisation distribution features two different topologies: a star singularity when $\mathbf{\hat{c}_h}=\mathbf{\hat{c}_-}$ and a lemon singularity  when $\mathbf{\hat{c}_h}=\mathbf{\hat{c}_+}$. The optical forces generated by FP beams have been studied before, therefore our analysis is restricted to the effect on the particle dynamics \cite{wang:opticalforces}. Figures \ref{fig:FPBexp1} and \ref{fig:FPBexp2} show the polarisation distribution, optical forces and particle dynamics for the two types of FP beam singularities. Notice that both FP beams have a flattop intensity which produces a central region where the scattering forces dominate the almost zero gradient force causing the particle to be expelled from the beam. We locate the particle near the optical axis to study the effect of the scattering forces. The resultant trajectories show that the dynamic of the nanoparticle depends on the type of singularity.
 
 We can see in figure \ref{fig:FPBexp1} that a FP beam with a star singularity produces spiral trajectories, similar to the ones studied by Berry et al. [14] by considering an idealized optical vortex of the form $U=(x+iy)^{\ell}$. This is due to the circular symmetry presented in all scattering forces and the null gradient force close to the singularity. However, FP beams with a lemon singularity have a curl force that breaks the rotational symmetry, and hence the dynamic does not generate spiral trajectories (see \ref{fig:FPBexp2}). 

\section{Off-axis superposition}

Let us consider a beam whose intensity profile is composed by two optical vortices with topological charges $|\ell_1|=|\ell_2|>0$ and are separated by a distance $2x_0$, i.e.
\begin{equation}
\mathbf{E}_{DFP}(x,y,z=0)= \frac{1}{\sqrt{2}} \, U_{\ell_1}(x+x_0,y)\,  \mathbf{\hat{c}_+} +\frac{1}{\sqrt{2}} \, U_{\ell_2}(x-x_0,y)\, \mathbf{\hat{c}_-}.
\label{eq:DFP}
\end{equation}

Since the polarisation singularities in this system have been studied in Lopez-Mago et al. \cite{dorilian:dynamics}, we restrict our analysis to the optical forces and dynamics with respect to the distance between the beams centroid. Analytic expressions for the optical forces for these fields are difficult to achieve, so numerical simulations were performed in order to analyse the complex dynamics.

\begin{figure}[H]
    \centering
    \includegraphics[width=12 cm]{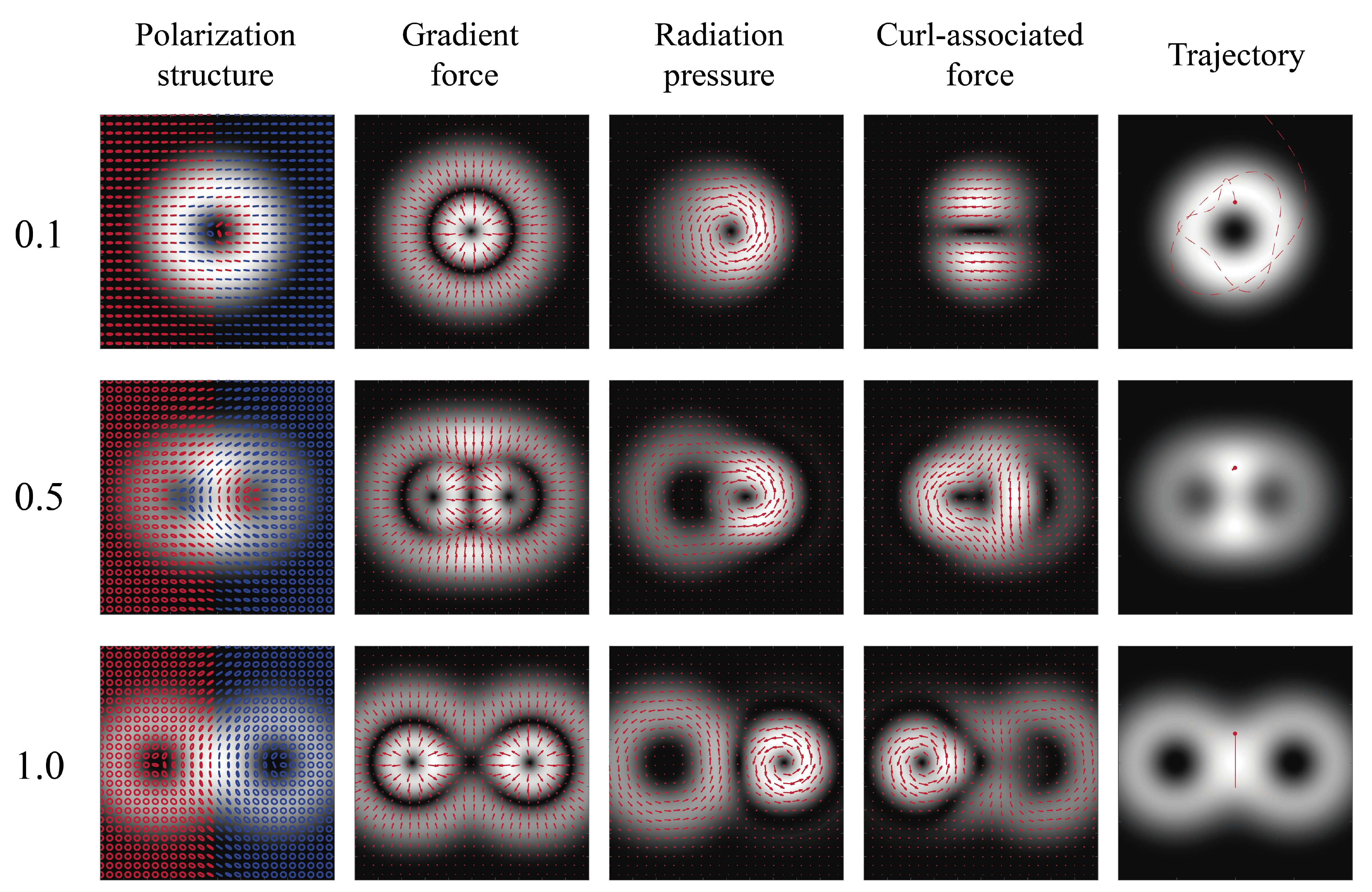}
    \caption{Off-axis composite optical vortices $\mathbf{C} = (\pi/4,\, \pi/4,\, \pi/2, \,0, 1, 1, x_0/w_0)$ for different values of $x_0/w_0$. When $x_0/w_0=0$, it degenerates to a Laguerre Gaussian beam with uniform horizontal polarisation.}
    \label{fig:DFP1}
\end{figure}

 First, let us consider the case with $\ell_1=\ell_2=1$, which for $x_{0} = 0$ corresponds to a uniform linear polarisation distribution. Figure \ref{fig:DFP1} shows the polarisation distribution, optical forces and trajectory followed by a nanoparticle for different values of $x_0$.  When the beams are almost completely aligned, the scattering forces are similar to the ones for a single optical vortices with linear polarisation, resulting in an open trajectory for the particle. As the separation distance $2x_0$ increases, complex and non symmetrical force fields are obtained which variate due the inhomogeneous distribution of $\mathbf{E}$ and $\mathbf
{H}$. Despite of the rupture of the symmetry, stable trapping and controlled oscillations are obtained for values of $x_0 \approx  w_0/\sqrt{2}$.
 
\begin{figure}[H]
    \centering
    \includegraphics[width=12 cm]{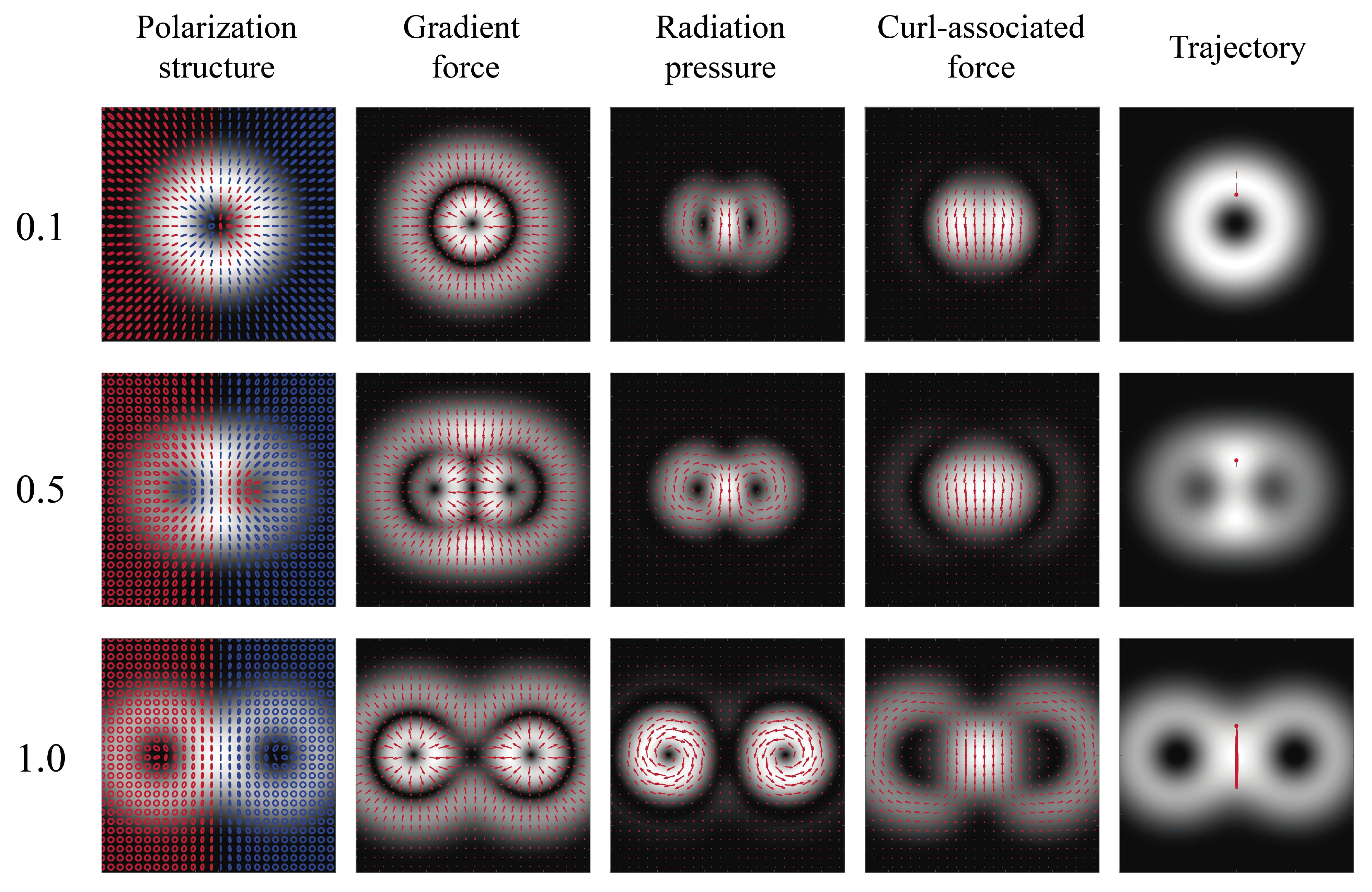}
    \caption{Off-axis composite optical vortices $\mathbf{C} = (\pi/4,\, \pi/4,\, \pi/2, \,0, 1, -1, x_0/w_0)$ for different values of $x_0/w_0$. When $x_0/w_0=0$, it degenerates to a radially polarised vector beam.}
    \label{fig:DFP2}
\end{figure}

\begin{figure}[H]
    \centering
    \includegraphics[width=12 cm]{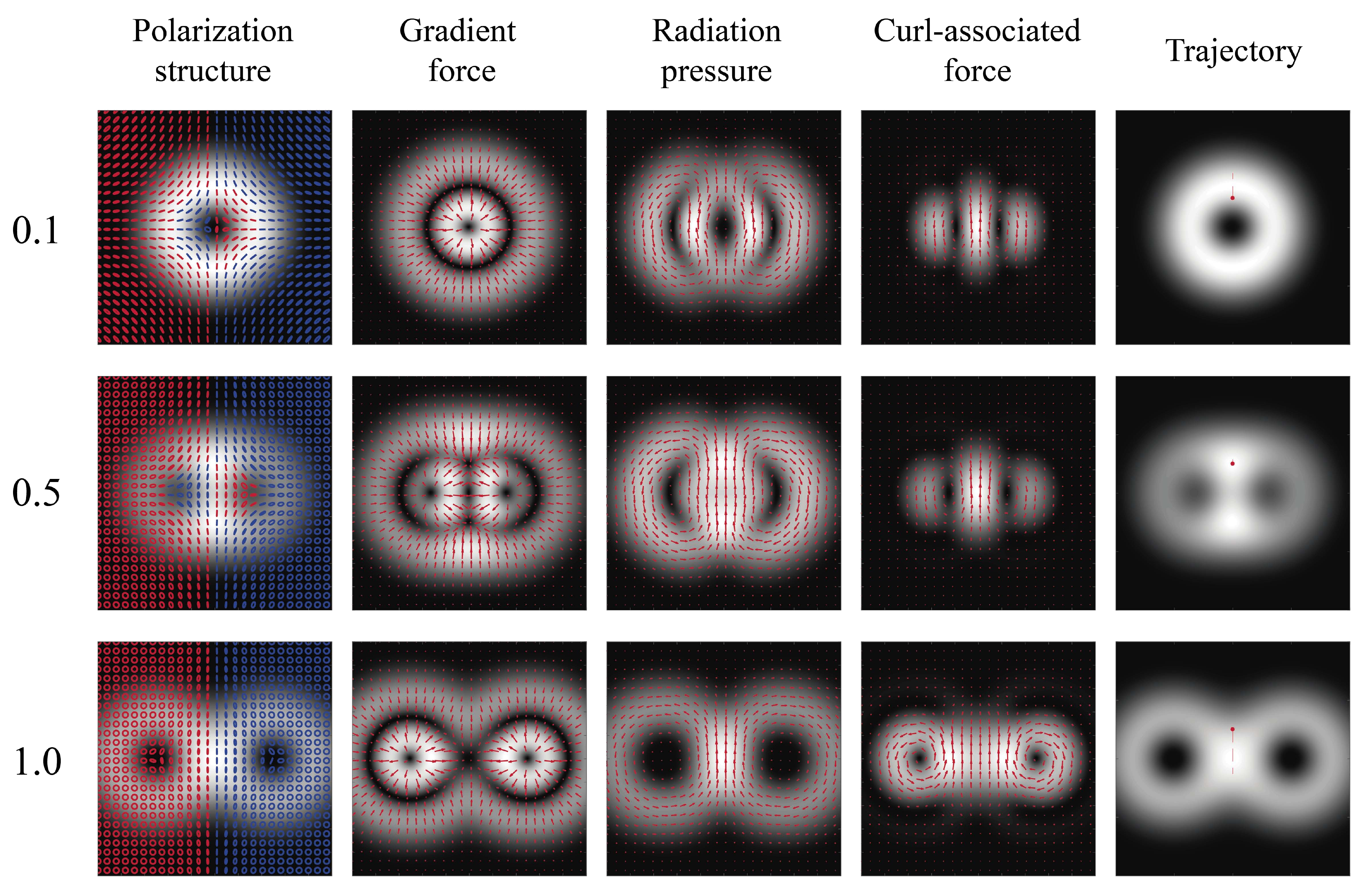}
    \caption{Off-axis composite optical vortices $\mathbf{C} = (\pi/4,\, \pi/4,\, \pi/2, \,0, -1, 1, x_0/w_0)$ for different values of $x_0/w_0$.When $x_0/w_0=0$, it degenerates to a hybrid polarised vector beam.}
    \label{fig:DFP3}
\end{figure}

 Similarly, we have explored the cases for $\ell_1=- \ell_2$. Some examples are shown in figures \ref{fig:DFP2} and \ref{fig:DFP3}. In the limit case when $x_0=0$, equation \ref{eq:DFP} becomes an special case of equation \ref{eq:CVB} with $\delta=0$. Similar to the previous case, for values of $x_0 \approx  r_{max}=w_0/\sqrt{2}$ trapping and controlled oscillations were obtained. Furthermore, since the optical forces are symmetric respect to the y axis, the superposition of the scattering forces generated by the optical vortices creates regions where both $\mathbf{F}_\mathrm{RP}$ and $\mathbf{F}_C$ have uniform and opposite directions. 

\section{Conclusions}
We have derived  analytical expressions for the optical forces that are generated by a single optical vortex. The importance of these expressions lies on the existence of scattering forces, $\mathbf{F}_\mathrm{RP}$ and $\mathbf{F}_\mathrm{C}$, are always present in homogeneously-polarised LG beams except for the particular case when $\beta=\ell=0$, which correspond to a fundamental Gaussian beam with uniform linear polarisation. In addition, independently of the polarisation state of the vortex and the initial condition for the particle, the resultant trajectory is a complex open orbit in which the particle is expelled due to the non-conservative scattering terms. 

Similarly, the optical forces generated by the CV beams and FP beams were analysed. The scattering forces in CV beams vanish independently on the topological charge and the relative phase between the circular components of the field $\delta$. Since only the gradient force is not zero, the particle oscillates respect to the closest intensity maximum describing a stable orbit. Meanwhile, FP beams have a flat top intensity which produces a region where the scattering forces dominate the almost zero gradient force and expel the particle from the beam. In addition, it has been shown that for identical initial conditions, the dynamics of the nanoparticle depend on the type of polarisation singularity. Interestingly, we found spiral trajectories for FP beams with a star singularity, where the scattering forces show circular symmetry.

We also presented a beam configuration with two optical vortices with topological charges $|\ell_1|=|\ell_2|>0$ and separated by a distance $2x_0$. For all the reviewed cases, trapping and controlled oscillations occurred for values of the separation distance $x_0 =  w_0/\sqrt{2}$.

In conclusion, we believe that these findings can have implications in experiments that involve trapping nanoparticles in high vacuum conditions. Furthermore, the study of spiral dynamics can now be explored using FP beams. 

\section{Acknowledgments}
Dorilian Lopez-Mago acknowledges support from Consejo Nacional de Ciencia y Tecnolog\'{i}a (CONACYT) through the grants: 257517, 280181, 293471, 295239, and APN2016-3140.

\end{document}